# Enhancing Curriculum Acceptance among Students with E-learning 2.0

Mr. Kamaljit I. Lakhtaria[1], Mr. Paresh Patel[2], Ms. Ankita Gandhi [3]
[1]Lecturer, MCA Dept., [2]Lecturer, IT. Dept., [3]Lecturer, C.E. Dept.
[1, 2]Atmiya Institute of Technology & Science, Rajkot, Gujarat
[3]Parul Institute of Engg. & Tech., Limda, Waghodia, Gujarat
[1] kamaljit.ilakhtaria@gmail.com, [2]mailtopareshpatel@yahoo.co.in, [3]anki.gandhi@gmail.com

**Abstract**

*E-learning; enhanced by communicating and interacting is becoming increasingly accepted and this puts Web 2.0 at the center of the new educational technologies. E-Learning 2.0 emerges as an innovative method of online learning for its incorporation of Web 2.0 tools. For any academic study, the curriculum provides overview of intact learning area. The Curriculum provides overview to content of the Subject. Many institutions place student interaction as a priority of their online curriculum design. It is proved that interaction has a great effect on the students' involvement in learning and acceptance of Curriculum. Students are accepting curriculum that is designed by teacher; whereas E-learning 2.0 enabled Curriculum management system allows student to involve in learning activities. It works as a stimulus and increases their dedication to the Curriculum. While Institute adapts E-Learning 2.0 as Learning Management System, it also provides Social Networking services and provides direct and transparent interaction between students and teachers. This view of the e-Learning 2.0 shifts its focus from LMS to the students, equipping them, with the means to become ever more autonomous, accepting them to make use of these means in solving problems on their own initiative. Curriculum usage will empower student involvement and enhancing E-learning 2.0 spreading. This paper, analyzing implementation E-learning 2.0 for Curriculum management and discusses Opportunities & Challenges for Curriculum over Web 2.0.*

## 1. Introduction

With Web 2.0 the Internet becomes a platform of fluid and continuous knowledge exchange. It is a dynamic social network of creation, sharing and consumption. Web 2.0 [1] technology creates innovative learning environments, allowing mentors, teachers & students to interact, collaborate, and create customized web enable e-learning experiences and offering an amazing potential to change the way learning, which will accepted by Institutes. E-Learning 2.0 emerges as an innovative method of online learning for its incorporation of Web 2.0 tools. Web 2.0 enable E-Learning 2.0 [2] isn't only about resorting to new technology, it is about a different approach to learning itself, it is a change in perception. E-learning 2.0 makes possible to minimize geographic and time restrictions and mobility concerns.

For any Academic course, always-common approach, Curriculum is designed by Teachers and followed by students. The Curriculum provides overview to content of the Subject. Mostly student follow curriculum as subject learning approach, mostly, student feedback / suggestion for subject learning and learning progress over subject never seems to reach to teachers. Lack of interactivity among students as well as students and teachers do not provide much acceptance and realization of effective curriculum for subject. The Web 2.0 enables E-learning Curriculum Management aims involvement of student as well as realization of effectiveness of curriculum to teachers.

## 2. Present Curriculum Management Tools

Many online curriculum management systems available, like Blackboard [3], Moodle [4], or WebCT. These all systems provide integrated solutions for faculty to post course content, assignments, and student evaluation. Rather no provision for student to post their reports, feedback or anything over there. So that present Curriculum Management tools likely to work only as document-centered, allowing instructors to post PowerPoint slides, Word and PDF files, and other course content for students to access.

In addition, course management systems allow students to log in to check grades, submit assignments, or take exams electronically. Teacher responsible for Subject content preparation and provide to student for download or access over these tools. ". Most common uses of these tools for teacher to provide learning material and student grades like to paste over Notice Board. Only advantage is student remotely access these all anytime at anywhere, no possibility to give a single over the same system.

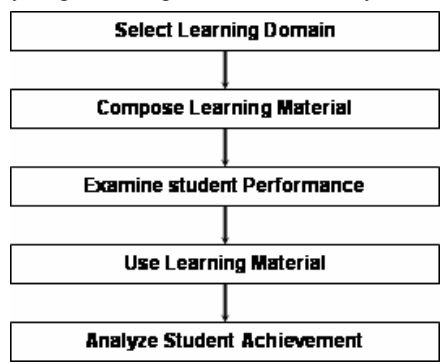

*Figure-1: Flow of the Institutional Learning Activities*

## 3. Prototyping Curriculum Management over Web 2.0

Traditional Classroom interaction among teachers and students is mostly one-way, even on Internet Teachers post Curriculum and student retrieve and read it. Today's course management systems are not being used to it's fullest potential. Vision of Web 2.0 based Curriculum Management system is to design "Course for student, course with students" not like earlier tools "built around the course, not the student". Most common uses are for faculty to distribute handouts and students to check grades. The role that the systems play most often is like that of an advanced photocopier, allowing faculty members to deliver materials to their students with greater ease.

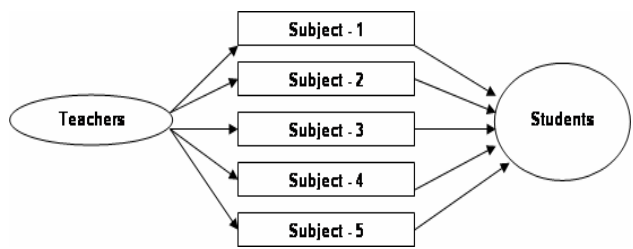

*Figure – 2: Traditional Learning System Direction (Uni-Direction)*

While the traditional curriculum best used for these common tasks, a next-generation curriculum over web 2.0 [5] must be centered on the student's learning, not the course's administration. Curriculum on Web 2.0 supplements the traditional curriculum and gives students a forum for collaborating and sharing their knowledge and understanding for the benefit of their classmates. Web 2.0, as a learning platform has its own concerns. Tracking user contributions may be difficult, and if the instructor provides too much structure, that could limit the E-learning 2.0's effectiveness. An instructor could structure and regulate interaction to such an extent that the E-learning 2.0 can effectively transform it into a stripped-down course management system. But doing so, risks diluting the special qualities that make E-learning 2.0worth using in the first place.

In web 2.0 enabled curriculums reciprocates and comments over web, representing curriculum. In Many Degrees students learn same curriculum but geographically resides away, while use of web 2.0 allows fetching response to each group and providing same to all other member students as well as in teachers also. This proposed way enhances not only student involvement but acceptance also.

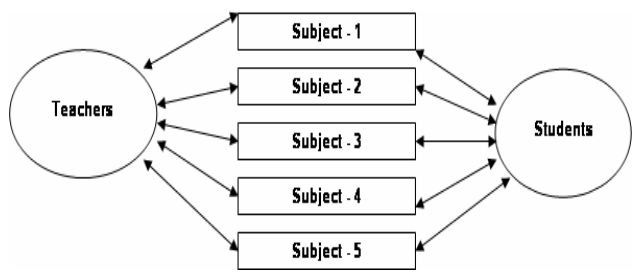

*Figure - 3: Web 2.0 enabled Learning System Direction (Active Participating – Bi-Direction)*

## 4. Architecture of Proposed System

The design of online curriculum comprises eight main aspects:

- Synopsis & Introduction of Subject;
- Outline of The Learning Goals;
- Assessment;
- Learning Material;
- Student Interaction;
- Technology;
- Student's Support
- Accessibility.

Proposed system needs following functions as key elements:

[1] Unified human interface
[2] Be able to publish, exchange, share and cooperate on information and knowledge among students and their instructor
[3] Be able to support idea generation activities in groups

### 4.1 Unification of human interface

The interfaces of the current web-based information systems are not unified and cause unnecessary confusion among students and instructors. Therefore the first improvement should be the unification of the human interfaces.

### 4.2 Insist student Involvement for Curriculum

Web 2.0 enable Curriculum management system needs design teams to develop their websites in order to publish, exchange, share, and cooperate on their design achievements. Design teams do not submit their achievements to the specified address anymore, but publish/open them to the public. Open Platform will enable students to think, work and publish their individual thought and expectation for subject wise curriculum. Teachers will also visit students' area over Curriculum portal for review their published data and give relevant and timely advice to them, as well as adapt suggestion out of them. Thus, the new Web 2.0 course management system will enable students and teachers to generate platform, which enables design solutions, publish exchange and share them, and cooperate on design projects.

Design activities need repeated involvement from both side students as well as teachers. For Institutes also likely that mentor or Principal also observe all such participation activities and outcome over teachers & students involvement. As per psychologically observation, repeated involvement over same process will definitely improve Curriculum Design, Management and Acceptance as well. Also empower student and teacher work together approach.

### 4.3 Installing a whiteboard to support idea generation activities in groups

One of the important design activities is to generate, visualize, structure and classify design solutions in groups. Mind Mapping Software, such as FreeMind [6], Mindomo [7], or MindMeister [8], will be helpful in generating design solutions by brainstorming. Therefore this paper proposes to design a white board into the new Web 2.0 Curriculum management system [11].

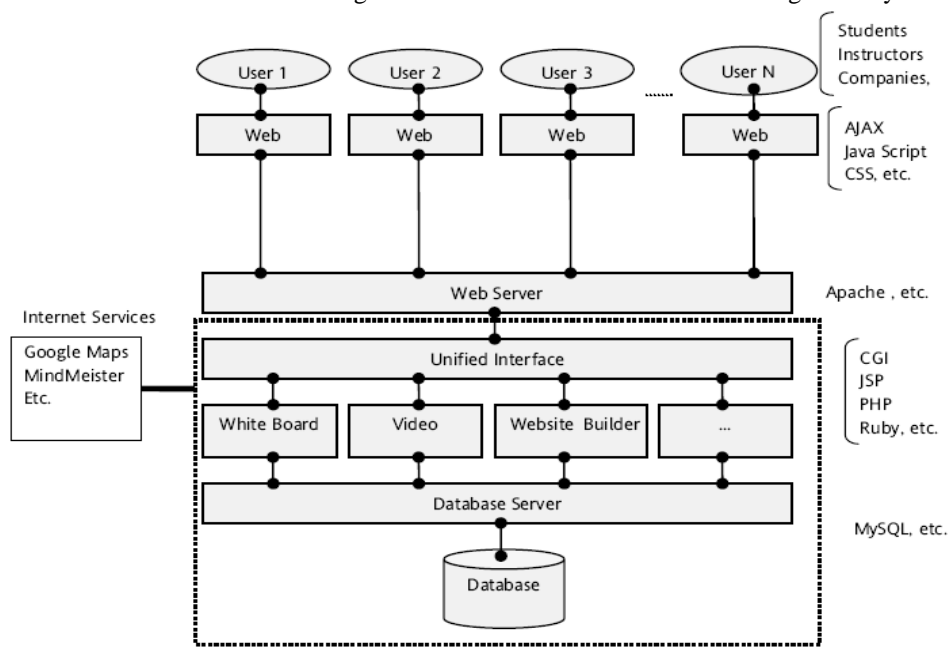

*Figure – 4: Overall Architecture of CMS*

### 5. Successful E-Learning Curriculum Management – Recommendations and Guidelines

The curriculum overview provides general information to the student about the content of the curriculum and its design and how it will be run. It is important to be simple so that the students understand the way the curriculum has been designed. Here, institutional information, sometimes neglected when thinking of designing a curriculum, should be included: regulations on privacy, plagiarism, and administrative and academic procedures. The outline of the learning goals will help the student to know what they are expected to learn and what they should be concentrating on. On the same note, the way in which the learners will be assessed is central to learning since it is the method used to track the learners' progress and evaluate how they are meeting the initial learning objectives. When preparing the material that will be used in the curriculum the instructor must ensure that they are adequate in the light of the established learning objectives, were assembled/produced by qualified and recognized sources and prepared with the purpose of supporting the peculiarities of distance learning.

Many institutions are, today, placing student interaction as a priority of their online curriculum design. It is proved that interaction has a great effect on the students' progress on the course. It works as a stimulus and increases their dedication to the course. This interaction has to be developed on three levels: learner-instructor, learner-learner and learner-content [9]. Technology has a role to play, although sometimes the integration of interactive tolls in online curriculum is criticized due to their lack of proven pedagogical significance. Technology enhances the learning experience and provides means and opportunity to students to interact among themselves and/or with the tutors. Technology will have influence on the students' learning and must be chosen effectively, it has to be accessible and easy to use, it should be either provided or easily obtainable and be accompanied of clear instructions of use. E-Learning 2.0 is based on flexibility and student's autonomy in the learning process, yet the preference given to the exclusive use of LMS, by the universities, does not support this prerogative.

With availability to social networks, their employment is essential in the sense that it works as a venue for the interaction of students among themselves and between them and the teachers. The students can share data and see one another's work, which not only grants access to their sources of information and improvement, but can generate debates. Use of a LMS discussion board is very different from using a blog, since the discussion emerges from an individual posting and a blog is more personal that a communal forum.

For teachers, it gives them opportunity to track the students' progress and if needed contribute to their work. Therefore, social networking has an important role in the development of closer relationships and it foments interaction between all actors in the learning process, while endowing them with the tools to reach outside the online course structure and connect with experts and peers anywhere in the globe. This view of the e-Learning 2.0 shifts its focus from LMS to the students, equipping them, with the means to become ever more autonomous, empowering them to make use of these means in solving problems on their own initiative. This method of organizing e- Learning 2.0 stimulates a learning experience that goes beyond the end of the course. The dynamism now embedded on the web is then transposed to e-Learning 2.0 in a way that allows for a more malleable education medium.

## 6. Conclusion

Education Technologies crosses new horizons, where growing expectations among Institute students to be able to access and manage their own course materials over the World Wide Web. Earlier, faculty would create web pages by hand for posting this information, While Internet technologies and access have matured over the past decade, course and learning management systems such as Blackboard have become the norm for distributing such materials. In today's Web 2.0 world, Curriculum published over web and enables students to contribute for enhance curriculum as well as subject learning process. Web 2.0 aiming to connect, collaborate and share people, with E-learning 2.0 Curriculum designing process not limiting to teachers only, but students also involve and student's involvement make them to accept the curriculum. For Teachers are final authorities to decide content of subject but students able to present their views on respective subject. Web 2.0 enable curriculum management will emerge as a tool that may complement or replace the use of traditional course management systems as a tool for disseminating course information. Building a course around the use of a Web 2.0 invites students to become involved in the process of creating course content and sharing their knowledge with their classmates.